\documentclass[bibyear]{aa}
\usepackage{ulem}
\usepackage{graphicx}
\usepackage{natbib}
\usepackage[hidelinks,colorlinks=true,linkcolor=red,citecolor=blue]{hyperref}
\usepackage{color}
\definecolor{DarkGreen}{rgb}{0.0, 0.5, 0.0}

\usepackage{txfonts}

\begin{document} 

\newcommand{\mjh}[1]{{\color{DarkGreen}#1}}

   \title{J0011+3217: A peculiar radio galaxy with a one-sided secondary lobe and misaligned giant primary lobes}
\titlerunning{J0011+3217: A peculiar radio galaxy}

\author{Shobha Kumari
          \inst{1}
          Sabyasachi Pal\inst{1}\thanks{E-mail: sabya.pal@gmail.com (SP)}
          Martin J. Hardcastle\inst{2}
          \and
          Maya A. Horton\inst{3}
}
 \authorrunning{Kumari et al.}
            
              \institute{Midnapore City College, Kuturia, Bhadutala, Paschim Medinipur, West Bengal, 721129, India
              \and
              Centre for Astrophysics Research, University of Hertfordshire, College Lane, Hatfield AL10 9AB, UK
              \and
              Astrophysics Group, Cavendish Laboratory, University of Cambridge, JJ Thomson Avenue, Cambridge CB3 0HE, UK
             }


\def \src{J0011+3217}
  \abstract
{From the LOFAR Two-metre Sky Survey second data release (LoTSS DR2) at 144 MHz, we identified a peculiar radio galaxy, J0011+3217. It has a large, one-sided diffuse secondary wing that stretches up to 0.85 Mpc (roughly 85\% of the size of the primary lobe). The linear size of the primary lobe of the galaxy is 0.99 Mpc. This peculiar source is a giant radio galaxy with a misaligned primary lobe. 
There is an optical galaxy 16 kpc (7 arcsec) from the host active galactic nucleus of J0011+3217. J0011+3217 has a radio luminosity of $1.65\times 10^{26}$ W Hz$^{-1}$ at 144 MHz with a spectral index of $-0.80$ between 144 and 607 MHz. J0011+3217 is located 1.2 Mpc from the centre of the Abell 7 cluster. The Abell 7 cluster has a redshift of 0.104 and a mass ($M_{500}$) of 3.71 $\times 10^{14}$ M$_\odot$. The cluster is associated with strong X-ray emission. We studied the X-ray emission around the cluster and from the region surrounding J0011+3217 using an \textit{XMM-Newton} image of J0011+3217, and we analysed the velocity structure and spatial distribution of galaxies in the cluster, showing that J0011+3217 inhabits an offset group of galaxies that are moving with respect to Abell 7. The off-axis distortion, or bending, of the primary lobe of J0011+3217 in the outer edges has a strong effect on the relative motion of the surrounding medium; this in turn causes the bending of the jets in the opposite direction of the cluster (like wide-angle tailed sources). We suggest that the morphology of J0011+3217 is influenced by ram pressure created by the Abell 7 cluster, highlighting the complex interactions between the source and the surrounding cluster environment.}
   
   \keywords{galaxies: ISM--galaxies: active--galaxies: clusters: intracluster medium               }
   
   \maketitle
%

\section{Introduction}
\label{sec:intro}
Radio galaxies consist of a pair of bipolar jets of supersonic plasma that can travel kiloparsec- to megaparsec-scale distances. An active galactic nucleus (AGN) resides at the centre of a typical radio galaxy. The majority of radio galaxies are not located in rich cluster environments, but those that often exhibit unique morphological traits because of their dense surroundings \citep{Ow97}. These unique morphological traits include extended radio lobes, powerful jets, and strong radio emission. The dense surroundings of rich cluster environments provide the necessary conditions for these galaxies to undergo intense interactions and mergers, leading to the formation of these distinct features.

The jets of radio galaxies are often distorted and have broken symmetries due to interactions with their surroundings, particularly in denser environments such as galaxy clusters.
These interactions lead to a wide range of morphological variations in the radio galaxies located in the inter-cluster medium (ICM). Variations in density inside the ICM, shock waves propagating through the ICM, and relative motion between the radio galaxy and the ICM cause distortions in the radio-emitting plasma and lead to complicated structures \citep{Be79, Jo17}. The non-uniform activity in the host AGN can also give rise to distortion in jets, either as a result of reorientations of the jets or due to variations in their jet properties.

Apart from normal radio galaxies, which exhibit a pair of lobes (primary lobes) oriented in two opposite directions, there is a sub-class of radio galaxies (X-shaped or Z-shaped winged radio galaxies) that exhibit an extra pair of lobes \citep[secondary wings or secondary lobes;][]{Ch07, Sa09, Go12, Ya19, Be20, Co20, Ku22, Bh22}. The origin of secondary wings in X-shaped radio galaxies (XRGs) is not well understood. Several models have been proposed to explain the extra wings in XRGs:
\begin{enumerate}
    \item The twin AGN model \citep{Ba80, Wo95}: The presence of a double AGN in the host galaxy produces secondary lobes in addition to the primary lobes. This model fails to describe the structure of many XRGs because, if true, secondary lobes should likewise have edge-brightened or edge-darkened structures, but no XRGs have hotspots in their secondary wings \citep{Go12, Ya19, Co20, Bh22}.
    \item The rapid jet reorientation model \citep{Me02}: The wings are the relics of a previous jet direction. In this picture, the jets have been rapidly reoriented due to the black hole--black hole merger. This model can easily accommodate secondary wings that are much larger than the primary lobes but requires invoking a physical process that, if it takes place at all, may be rare.
    \item Backflow plasma model (a widely accepted scenario): The secondary wings are caused by the synchrotron plasma, which backflows after a deflection by an asymmetric thermal halo encircling the optical core \citep{Le84, Ca02}. The Fanaroff--Riley type II (FRII) sources most commonly show an X-shaped structure \citep{Ca02, Sa09, Go12, Co20, Be20, Bh22} that follows the backflow model. A few FRI sources were recently found to also possess an X-shaped structure \citep{Sa09, Go12, Co20, Ya19, Bh22}. 
    \item Slow jet reorientation (precession) model: As noted by \cite{Ho20}, simulations of slowly precessing jets can generate X-shaped sources even in symmetrical environments as the jet moves its termination point towards the edge of a pre-existing lobe.
    \end{enumerate}

\begin{figure}
\includegraphics[width=9cm,origin=c]{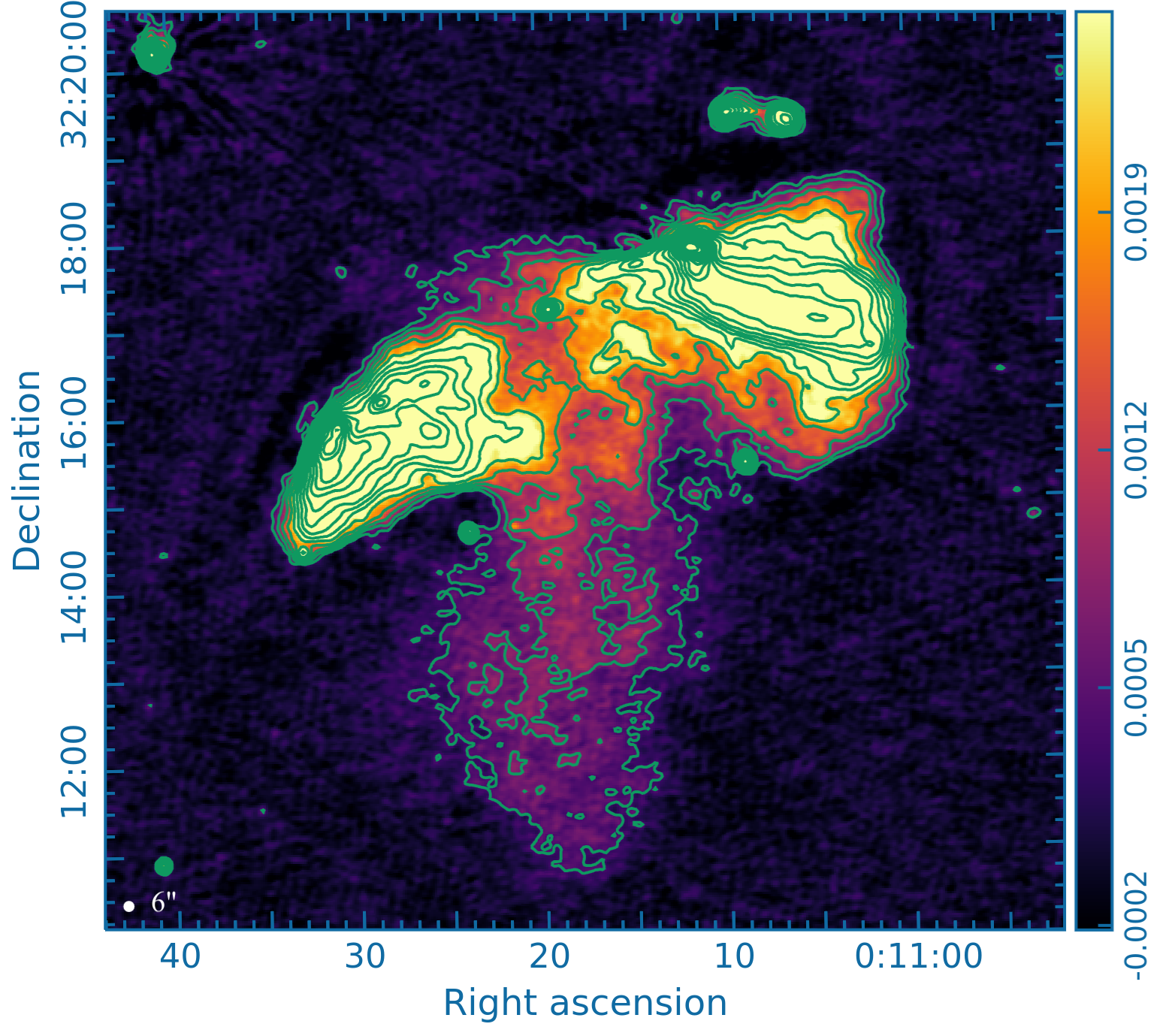}
\caption{LOFAR image of \src{} at 144 MHz. The contour levels are at 0.23, 0.45, 0.75, 1.17, 1.75, 2.57, 3.72, 5.33, 7.58, 10.7, 15.1, 21.3, 30.0, 42.1, and 59.0 mJy beam$^{-1}$.  The lowest contour level of \src{} is at 3$\sigma$, where $\sigma=78$ $\mu$Jy beam$^{-1}$ is the mean RMS around the source. The white circle in the left corner of the image represents the synthesised beam of \src{} with 6\arcsec$\times$6\arcsec resolution.}
\label{fig:LOFAR}
\end{figure}
In this paper we report on a peculiar giant radio galaxy that has a one-sided secondary wing. In Sect. \ref{sec:data} we explain the data reduction and analysis. In Sect. \ref{sec:result} we present our results and discuss them in Sect. \ref{sec:discuss}. Section \ref{sec:conclusion} summarises this paper.
We used the following cosmology parameters, which are the final full-mission \textit{Planck} measurements of the cosmic microwave background anisotropies: $H_0$ = 67.4 km s$^{-1}$ Mpc$^{-1}$, $\Omega_{vac}$ = 0.685, and $\Omega_m$ = 0.315 \citep{Ag20}.

\begin{figure}
\includegraphics[width=8.5cm, origin=c]{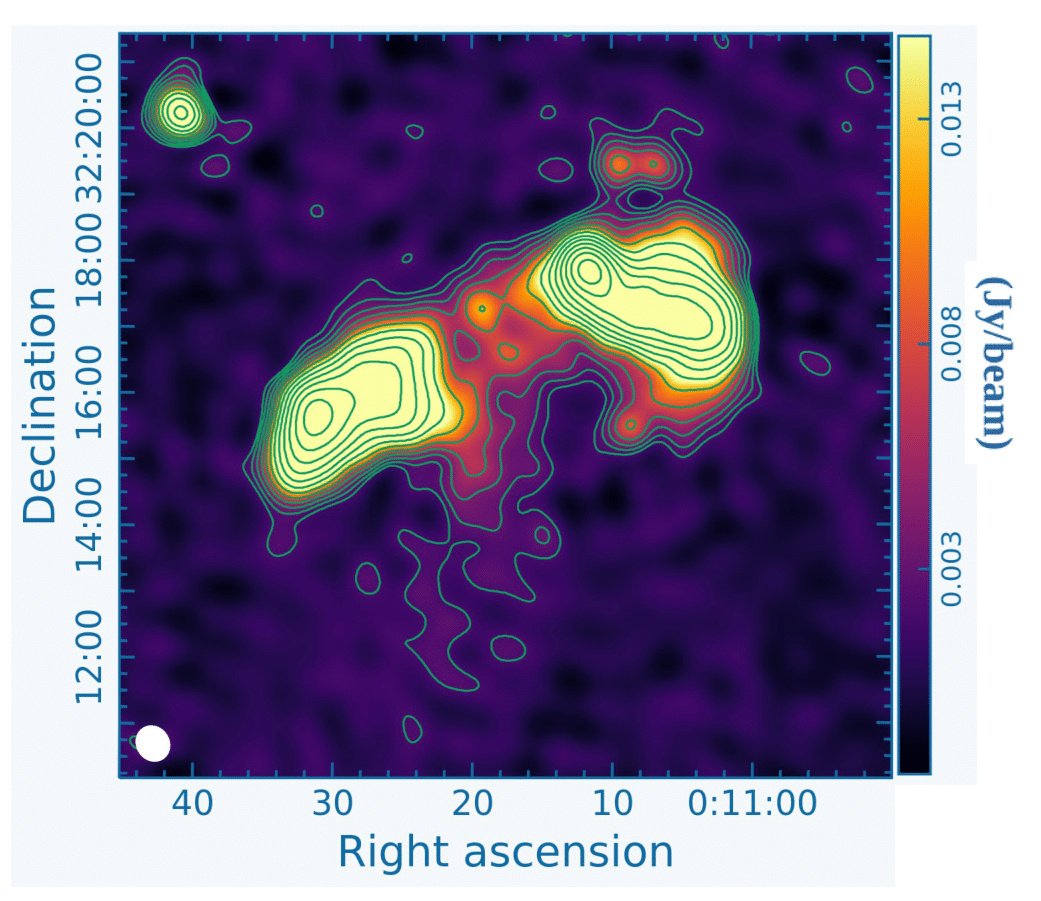}
\caption{GMRT image of \src{} at 607 MHz with a 5k$\lambda$ UV distance cutoff. The contour levels are at  1.02, 1.82, 2.94, 4.51, 6.70, 9.78, 14.1, 20.1, 28.5, 40.3, 56.9, 80, 110, and 160 mJy beam$^{-1}$. The lowest contour level is 3$\sigma$, where $\sigma=0.2$ mJy beam$^{-1}$ is the mean RMS around the source. The synthesised beam (resolution: 32.3\arcsec$\times$ 28.7\arcsec, position angle: 32.5$^\circ$) is denoted with a filled white ellipse in the lower-left corner of the image.}
\label{fig:GMRT_5k}
\end{figure}

\section{Data reduction and analysis}
\label{sec:data}
\subsection{LOFAR survey data}
\label{subsec:lofar}
The peculiar radio galaxy J0011+3217\footnote{This object was first catalogued as a bright radio source in the B2 survey, and so is best known as B2 0008+32 \citep{Co70}. In the LoTSS DR2 optical identification catalogue \citep{Hardcastle23}, this object has the standard LOFAR identifier ILTJ001121.28+321638.8. Throughout the current article, we refer to the object by the shorter name J0011+3217 for convenience.} was identified using the International LOw-Frequency ARray (LOFAR) Two-metre Sky Survey second data release \citep[LoTSS DR2;][]{Sh22}. LOFAR is currently the largest ground-based radio telescope operating in the low-frequency range of 120--168 MHz. The survey images were developed using a fully automated, direction-dependent calibration and imaging pipeline. The survey consists of mosaics of several 8-hour LOFAR observations. The survey has a median RMS sensitivity of 83 $\mu$Jy beam$^{-1}$ with 6\arcsec\ resolution and operates at a central frequency of 144 MHz.
\begin{figure*}
\centerline{
\includegraphics[width=18cm,origin=c]{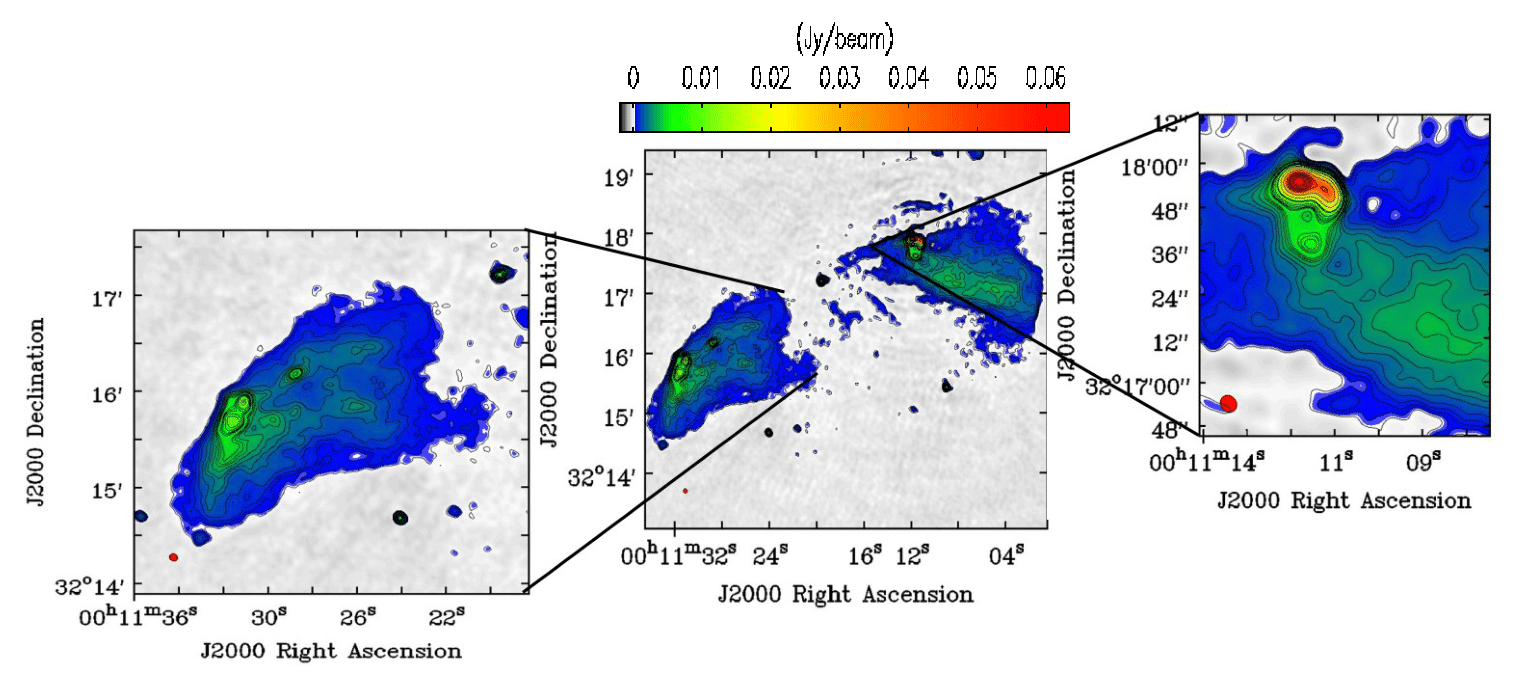}
}
\caption{GMRT image of J0011+3217 at 607 MHz using no UV cutoff. Zoomed-in versions of the eastern lobe and the misaligned western lobe are shown in separate panels. The lowest contour level is 3$\sigma$, where $\sigma$ is the RMS noise of the 607 MHz GMRT image. The RMS noise of the image is 40 $\mu$Jy. The contour levels are at 3$\sigma\times$(1, 1.4, 2, 2.8, 4, 5.7, 8, 11, 16, 23, 32, 45, 64, and 75). The synthesised beam is 5.0\arcsec$\times$4.4\arcsec  (position angle 46.6$^\circ$) and is denoted with the filled red circle in the lower-left corner of the image.}
\label{fig:GMRT_full}
\end{figure*}
\subsection{GMRT data analysis}
\label{subsec:gmrt}
To study the nature of the source at a relatively high frequency, we used Giant Meterwave Radio Telescope \citep[GMRT;][]{Sw91} archival data\footnote{https://naps.ncra.tifr.res.in/goa/data/search} (proposal code:29\_058)  with a central frequency of 607 MHz.  The source was observed for 7h using multiple scans. 3C 48 was used as the flux calibrator and observed for 15 min, whereas 0029+349 was used as the phase calibrator and observed for 1.8 hours in various scans. The data were recorded in 256 frequency channels. We used the Common Astronomy Software Applications ({\tt CASA}) package for data analysis \citep{Mc07}. Standard steps of flagging (removal of bad data), calibration of complex gains, and bandpass are carried out.
The flux density scale of \citet{Pe17} is used for absolute flux calibration. Several rounds of phase-only self-calibration followed by amplitude and phase self-calibration are carried out to improve sensitivity. We imaged the final visibilities using the Briggs weighting scheme  \citep{Br95}, with robust = 0. To observe the diffuse emission of the source at 607 MHz, we used a UV distance cut of 5k$\lambda$. 
J0011+3217 is located 9.6 arcmin from the centre of the field. A primary beam correction was applied to the image during the analysis.

\subsection{XMM-Newton data analysis}
\label{subsubsec:XMM}
\textit{XMM-Newton} \citep{Ja01, de01} observation (obs id: 0827021201) was carried out centred on the nearby cluster Abell 7 (see Sect. \ref{sec:cluster}) with field centre of $\alpha$:00h11m41.85s; $\delta$: +32$^{\circ}$24$'$33.3$''$ on 20-06-2018. The two Multi-Object Spectrometer cameras  (MOS1 and MOS2) and the European Photon Imaging Camera (EPIC pn) on board \textit{XMM-Newton} were operated in the nominal full-frame mode. The exposure time for MOS1, MOS2, and pn data were 29752s, 31192s, and 24190s, respectively. We used \textit{XMM-Newton} Science Analysis System (SAS) v18.0.0 \citep{Ga04} for \textit{XMM-Newton} European Photon Imaging Camera (EPIC) data reduction. All the obtained data were reprocessed based on the latest current calibration files. To obtain the photon event lists, we employed tasks {\tt EMPROC} and {\tt EPPROC} on the MOS and pn event files. The out-of-time pn event files are produced by the task {\tt EPPROC}. To apply the standard filter to the event files and create the image, we used task {\tt EVSELECT} to select single, double, triple, and quadruple events (PATTERN$\leq$12) for MOS and single and double events (PATTERN$\leq$4) for pn, with an energy range from 0.2 to 12 keV. FLAG was set to zero to reject events that were close to the charge-coupled device (CCD) gap and that contained bad pixels. In addition, we also imposed the criterion of \#XMMEA\_EM and \#XMMEA\_EP for MOS and PN, respectively, in order to filter out artefact events. We then filtered for background flares in the standard manner, which reduces the pn exposure time to 20.4 ks. The response matrix file (rmf) and the ancillary response file (arf) were created using the SAS tools {\tt rmfgen} and {\tt arfgen}, respectively. For spectral extraction, spectra were grouped to a minimum of 100 counts per bin before background subtraction using the task {\tt GRPPHA} in {\tt FTOOLS} in order to facilitate standard $\chi^2$ fitting techniques.

We also inspected an archival short {\it Chandra} Advanced CCD Imaging Spectrometer (ACIS-I) observation of Abell 7, observation ID 15157, but J0011+3217 lies outside the field of view of this observation.

\section{Results}
\label{sec:result}
We identified a peculiar radio galaxy, J0011+3217, with a large one-sided diffuse secondary wing from the LoTSS DR2 images. The secondary wing originates from the core of the galaxy and extends 0.85 Mpc given the redshift inferred for the host galaxy (more on this in Sect. \ref{subsec:optical-host}). Figure \ref{fig:LOFAR} presents the LOFAR image of J0011+3217 at 144 MHz with 6\arcsec$\times$6\arcsec\ resolution. We also checked the 20\arcsec\  resolution LoTSS DR2 image at 144 MHz and found no additional diffuse emission features. The continuum emission from J0011+3217 at both resolutions has the same features as a large one-sided diffuse wing. An interesting feature of this peculiar radio galaxy J0011+3217 is that the lobes of the radio galaxy are misaligned. The lobes are bent in a common direction with a misalignment angle $\sim$50$^{\circ}$ (see Figs. \ref{fig:LOFAR} and \ref{fig:GMRT_5k}). The misalignment angle was measured by drawing two straight lines parallel to two primary radio jets of J0011+3217 and taking the angle that is within the range 0 to 90 degrees.

\begin{figure*}
\vbox{
\centerline{
\includegraphics[width=14cm,origin=c]{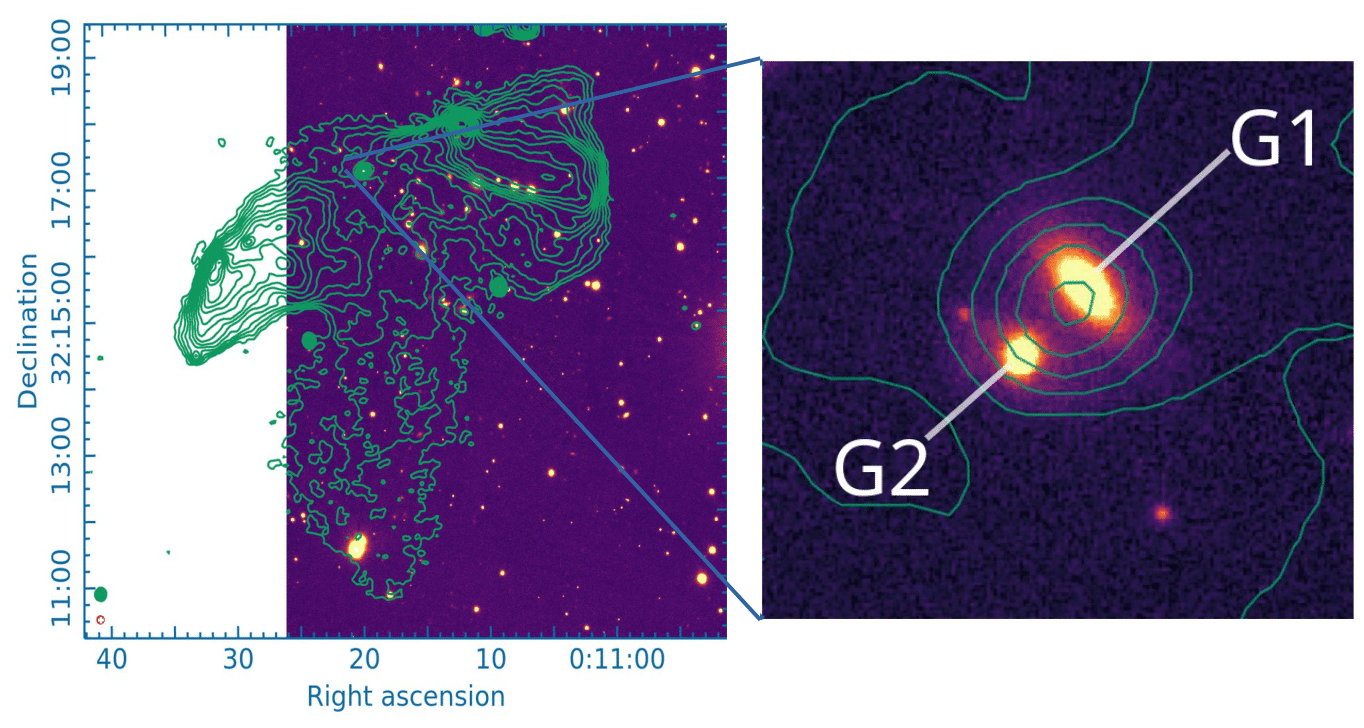}
}
}
\caption{LoTSS DR2 image of the peculiar radio galaxy J0011+3217 in contours (levels: 0.23, 0.45, 0.75, 1.17, 1.75, 2.57, 3.72, 5.33, 7.58, 10.7, 15.1, 21.3, 30.0, 42.1, and 59.0 mJy beam$^{-1}$) overlaid on an image from the Pan-STARRS1 optical r band. The right panel is a zoomed-in view of the optical galaxies G1 and G2. The lowest contour level of \src{} is at 3$\sigma$, where $\sigma=78~\mu$Jy beam$^{-1}$ is the RMS of the field. The synthesised beam of the LoTSS DR2 image (marked by the red circle in the lower-left corner of the image) has 6\arcsec$\times$6\arcsec resolution. }
\label{fig:optical}
\end{figure*}

The source has two hotspots on either side of the core, which are embedded in the radio lobes, and so is clearly an FRII radio galaxy. To study the detailed morphological nature and confirm the diffuse secondary wing of this source, we also present a GMRT image of this source at 607 MHz. In the full-resolution (with synthesised beam 5.0\arcsec$\times$4.4\arcsec, position angle 46.6$^\circ$) image of \src{} at 607 MHz, the diffuse secondary lobe is not visible (see Fig. \ref{fig:GMRT_full}). To study the diffuse emission of this source, we set a UV distance cutoff of 5k$\lambda$ in the CASA {\tt tclean} task. Figure \ref{fig:GMRT_5k} presents a GMRT 607 MHz image with a diffuse secondary wing at a UV distance cutoff of 5k$\lambda$. 

\subsection{Optical and IR counterpart: Interacting pair of optical galaxies}
\label{subsec:optical-host}
\begin{figure*}
\vbox{
\centerline{
\includegraphics[width=14cm,origin=c]{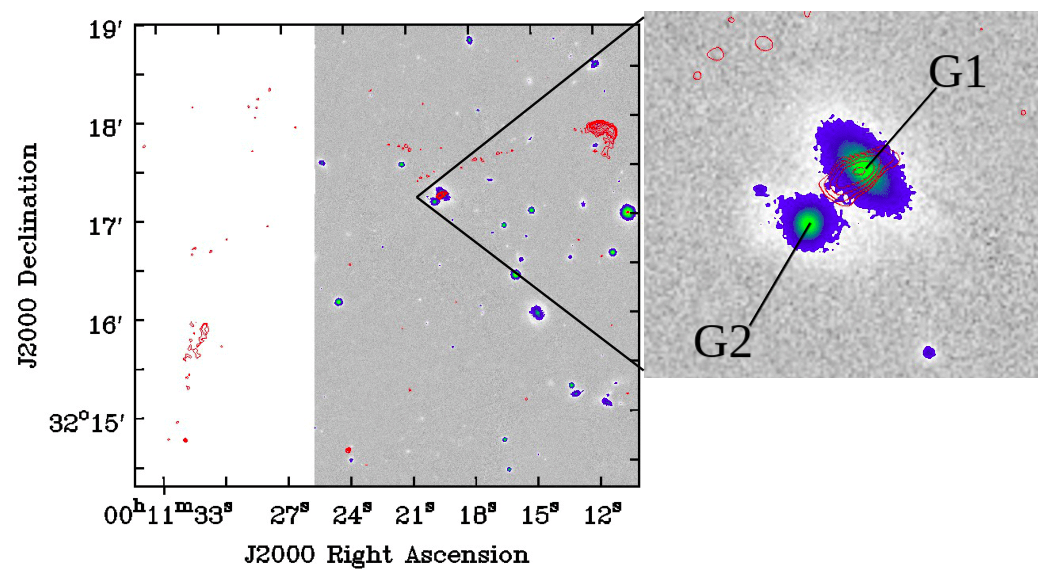}
}
}
\caption{High-resolution VLASS image of the peculiar radio galaxy J0011+3217 (red contours) overlaid with Pan-STARRS1 optical r band. The right panel is a zoomed-in view of the optical core where galaxy G1 coincides with the radio core. The synthesised beam (marked by the filled red elliptical in the lower-left corner of the image) of the VLASS image is 2.6\arcsec$\times$2.3\arcsec; the position angle is --79.2$^{\circ}$.}
\label{fig:VLASS+optical}
\end{figure*}

We carry out an independent optical identification of the optical counterpart of J0011+3217 using the Panoramic Survey Telescope and Rapid Response System \citep[Pan-STARRS1;][]{Ch16}. Figure \ref{fig:optical} shows an optical r-band Pan-STARRS1 \citep{Ch16} image of J0011+3217 overlaid with a LOFAR image of J0011+3217 at 144 MHz. To identify the radio core of J0011+3217, we used the Very Large Array (VLA) Sky Survey (VLASS) \citep{La20} image of J0011+3217 at 3 GHz with 2.5\arcsec resolution, which we show in Fig. \ref{fig:VLASS+optical}. The radio core of the VLASS image of J0011+3217 coincides with the optical counterpart (SDSS J001119.35+321713.8 or WISEA J001119.34+321713.8) denoted by G1 in the right panel of Fig. \ref{fig:VLASS+optical}. We conclude that G1 is the optical and IR host galaxy of J0011+3217, as also reported in the LoTSS DR2 optical ID catalogue by \cite{Hardcastle23}. Due to the low resolution in the LOFAR image in Fig. \ref{fig:optical}, an offset between the radio and optical peak can be seen whereas the high-resolution VLASS image in Fig. \ref{fig:VLASS+optical} does not show any offset between the radio and optical peak. The optical galaxy G1 has a spectroscopic redshift of 0.107144 \citep{Ahu20}. This corresponds to a physical scale of 2.031 kpc arcsec$^{-1}$ for our adopted cosmology.

The optical or IR galaxy (SDSS J001119.75+321709.0 or WISEA J001119.75+321709.0), denoted by G2 in the zoomed right panel of Figs. \ref{fig:optical} and \ref{fig:VLASS+optical} has a spectroscopic redshift of 0.105552. The projected linear distance between the optical galaxies G1 and G2 is 16 kpc (7 arcsec).

\subsection{Cluster and group environment}
\label{sec:cluster}

J0011+3217 resides at a distance of 9.9 arcmin (1.2 Mpc projected distance) away from the centre of the cluster Abell 7, also known as {\it Planck} cluster PSZ2 G113.29--29.69. 
The mass of the Abell 7 cluster (M$_{500}$) within the $r_{500}$ radius is $(3.71\pm0.27)\times 10^{14}$ M$_{\odot}$, with $r_{500} =1060\pm25$ kpc \citep{Bo22}. The cluster lies at a very similar mean redshift ($z=0.104$, \citealt{Rozo15}) to the host galaxy G1 and its companion G2, so it seems very likely that the radio galaxy is physically associated with the cluster.

\begin{figure*}
\includegraphics[width=\linewidth]{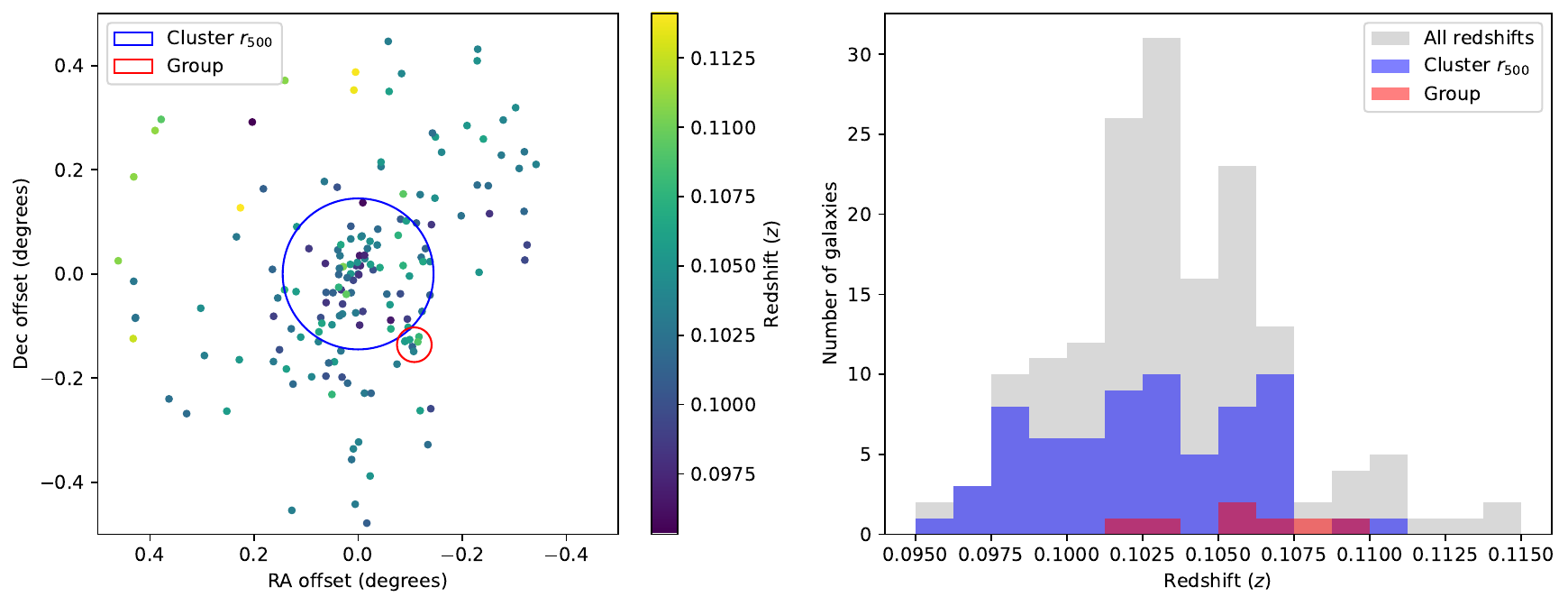}
\caption{Distribution of the redshifts of the cluster and the nearby galaxies. Left: Sky distribution of galaxies in the cluster Abell 7 with $0.095 < z < 0.115$. The large blue circle shows the $r_{500}$ for the cluster as described in the text, and the small red circle indicates a 2 arcmin ($\sim 250$ kpc) radius around the host of J0011+3217. Note that a bright star at the bottom right of the plot may prevent the detection of galaxies in this area. Right: Histogram of the redshifts in the left panel, with galaxies within $r_{500}$ highlighted in blue and those close to the radio source in red. If some or all of the galaxies close to the radio source form a group, the group has a systematic velocity offset with respect to the cluster.}
\label{fig:A7_galaxies}
\end{figure*}

We used the NASA Extragalactic Database (NED) to search for known galaxies within 30 arcmin (3.6 Mpc) of the centre of Abell 7, and as expected, the distribution of their spectroscopic redshifts from the Sloan Digital Sky Survey \citep[SDSS;][]{Ahu20} is strongly peaked around $z \sim 0.105$, with 162 galaxies with redshifts having $0.095 < z < 0.115$, of which 62 lie within $r_{500}$ in projection. The bulk of the galaxies in this redshift slice are concentrated around the centre of Abell 7 but there is a clear overdensity (see Fig. \ref{fig:A7_galaxies}) of 6-8 bright galaxies around the position of J011+3217 and its host galaxy G1. Taking the mean redshifts of the nearest 7 of these galaxies and comparing them to the 62 objects within $r_{\rm 500}$, there is a radial velocity offset of the order $(1000 \pm 300)$ km s$^{-1}$, which suggests that the galaxies around G1 may be a group interacting with the cluster.

\begin{figure}
\includegraphics[width=9cm,origin=c]{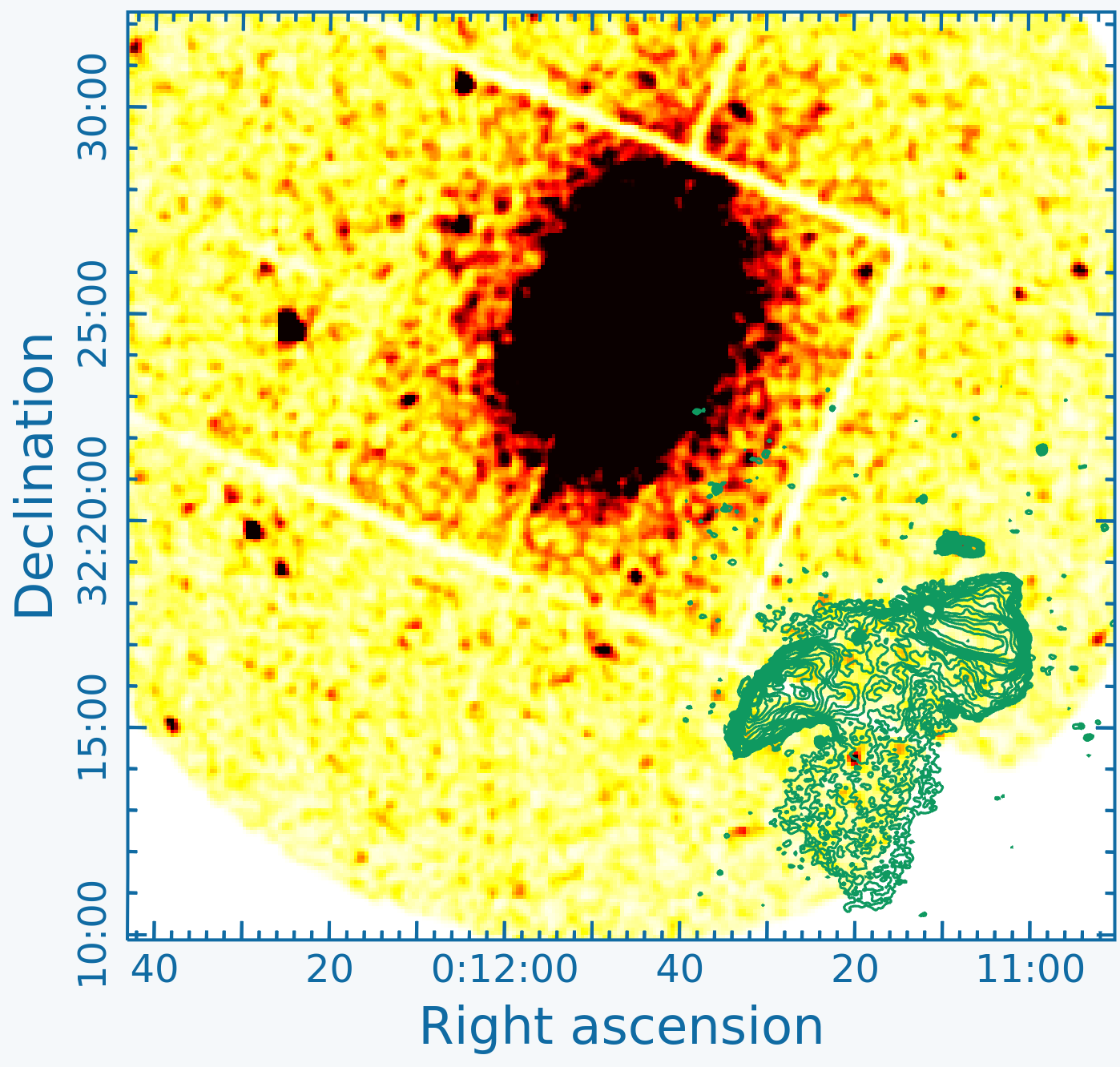}
\caption{Smoothed \textit{XMM-Newton} MOS image of Abell 7 in the energy range 2--10 keV, overlaid with the LOFAR image of \src{} at 144 MHz (contour levels: 0.23, 0.45, 0.75, 1.17, 1.75, 2.57, 3.72, 5.33, 7.58, 10.7, 15.1, 21.3, 30.0, 42.1, and 59.0 mJy beam$^{-1}$).}
\label{fig:X-ray}
\end{figure}

We searched for X-ray emission around the cluster and from the large structure of J0011+3217 and detected strong X-ray emission from the centre of Abell 7. The overlaid X-ray image at the energy range 2--10 keV with the LOFAR image (6\arcsec) at 144 MHz of J0011+3217 is presented in Fig. \ref{fig:X-ray}. There is no evidence of either extended or compact X-ray emission coincident with the radio galaxy itself or the group that may surround it. This could imply that interactions with the cluster have partially or wholly stripped away the hot-gas environment of the group.

We extracted a spectrum from the central 5 arcmin of the {\it XMM-Newton} observation, which allowed us to measure\footnote{We fitted using {\sc xspec}, assuming a Galactic $N_H$ of $4.40 \times 10^{20}$ cm$^{-2}$ and taking an annular background subtraction region of 5--5.5 arcmin around the cluster. The metal abundance, a free parameter of the fit, is $0.31 \pm 0.03$ solar.} a cluster temperature of $4.56 \pm 0.095$ keV for the central regions of the cluster, consistent with the temperature in the outer regions of $4.05 \pm 0.07$ keV measured by \cite{Iqbal23} as part of their temperature profile analysis. This temperature, which is also consistent with the temperature-$M_{500}$ relation of \cite{Arnaud05}, implies a sound speed for the central regions of the cluster around 1000 km s$^{-1}$, so on the rough velocity offset estimate derived above, the motion of the group hosting the radio galaxy is likely to be transonic or supersonic with respect to the cluster gas, particularly if the temperature is lower beyond $r_{500}$.

\subsection{Flux density measurements and spectral index map}
\label{subsec:spectral}
The measured total integrated flux densities of J0011+3217 at 144 MHz (including diffuse emission of LOFAR image at 6\arcsec$\times$6\arcsec resolution) and 607 MHz (GMRT image at UV distance cut of 5k$\lambda$ including diffuse emission) are 6.30$\pm0.31$ Jy and 1.55$\pm0.08$ Jy, respectively. The error in flux density measurements for 144 MHz LOFAR images \citep{Sh22} is $\sim$5\%.

The error associated with our measured flux density at 607 MHz GMRT image is calculated using Eq. \ref{eq:error} \citep{phoenix}:
\begin{equation}
\sigma_I=I \sqrt{2.5 \frac{\sigma^2}{I^2} + 0.05^2}
\label{eq:error}
,\end{equation}
where $\sigma_I$ is the total error on the integrated flux density, $\sigma$ is the RMS error in the image, and $I$ is the total integrated flux density of \src{}.  
We assume 0.05 (5\%) as the calibration error of GMRT \citep{Sw91}. 

Using the measured flux density, we calculated the two-point spectral index of this peculiar radio galaxy between 144 MHz and 607 MHz. We included the diffuse emission; the GMRT image at 607 MHz with UV distance cut of 5k$\lambda$ and measured the spectral index as --0.97$\pm0.05$ (using $S_{\nu} \propto \nu^{\alpha}$). Here, $S_{\nu}$ is the radiative flux density at a given frequency $\nu$ and $\alpha$ is the spectral index. 

\begin{figure*}
\vbox{
\centerline{
\includegraphics[width=9.5cm,origin=c]{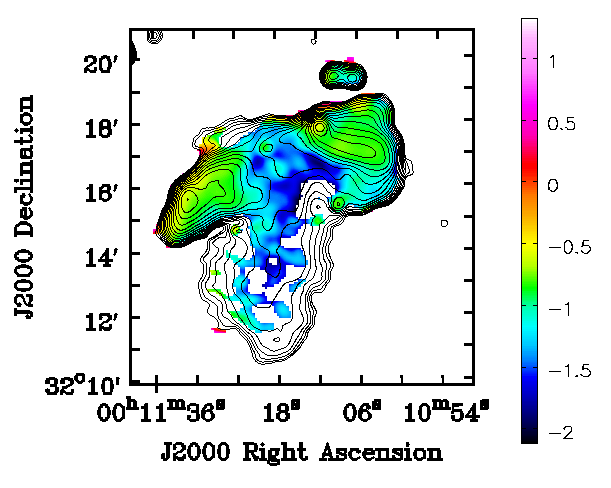}
\includegraphics[width=9.5cm, origin=c]{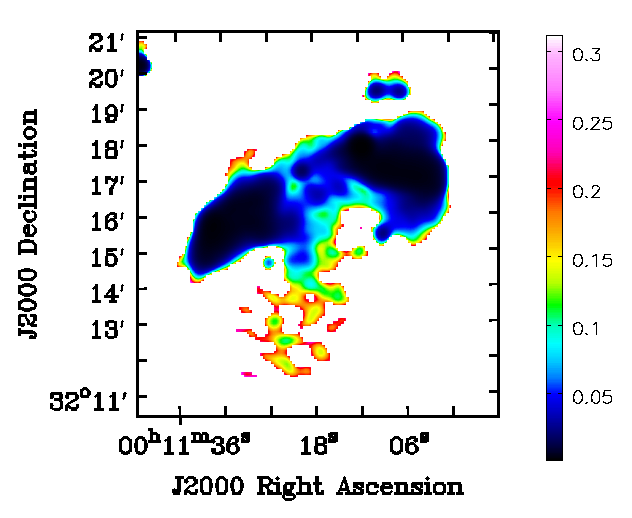}
}
}
\caption{Spectral index map of \src{} between 144 MHz and 607 MHz of the peculiar radio galaxy J0011+3217. Left: A LoTSS DR2 image at 144 MHz (with 20\arcsec$\times$20\arcsec resolution) is overlaid in contours (levels: 0.23, 0.45, 0.75, 1.17, 1.75, 2.57, 3.72, 5.33, 7.58, 10.7, 15.1, 21.3, 30.0, 42.1, and 59.0 mJy beam$^{-1}$) with the spectral index map of \src{}. Right: Spectral index error map of \src{} between 144 MHz and 607 MHz .}
\label{fig:spec_map}
\end{figure*}

The left panel of Fig. \ref{fig:spec_map} presents the spectral index map of \src{} between 144 and 607 MHz, and the right panel of Fig. \ref{fig:spec_map} presents the corresponding error in the spectral index map of \src{}. The contour plot in Fig. \ref{fig:spec_map} shows the LOFAR image of \src{} at 144 MHz with 20\arcsec$\times$20\arcsec. The typical error in the spectral index map is $\sim$0.05 (see the right panel of Fig. \ref{fig:spec_map}). The error is uniform, except at the edge of the radio emission, where it is up to 0.3. For the spectral index map, we used the LOFAR image at 144 MHz with 20\arcsec$\times$20\arcsec resolution and the GMRT image at 607 MHz. The full resolution GMRT image (5.0\arcsec$\times$4.4\arcsec; position angle 46.6$^\circ$) does not detect one-sided extended diffuse emission of \src{} as observed in the LOFAR image. We therefore use a uv-taper of 22\arcsec in the GMRT 607 MHz image to detect the diffuse secondary wing. The tapered GMRT image captures the one-sided extended diffuse emission with a resolution of $\sim$21\arcsec. To create the spectral index map we compare this with the LOFAR image at 20\arcsec, as this has a very similar resolution and is expected not to miss any extended diffuse emission. 
We convolve both images (LOFAR and GMRT images) to a resolution of 21\arcsec$\times$21\arcsec. To create the sub-images of the same size (pixel to pixel), we used the {\sc aips} task {\tt OHGEO}. After that, we used the task {\tt COMB} with OPCODE `{\tt SPIX}' to produce the spectral index map and spectral index error map of J0011+3217. Here we use 3$\sigma$ clip for both images. The spectral index at the hotspots of both sides of the structure is $\sim -0.60$. The spectral index in the right or western lobe varies from $-0.5$ to $-1.2$ whereas in the left or eastern lobe varies from $-0.62$ to $-0.89$ (see the left panel of Fig. \ref{fig:spec_map}). The western lobe is steeper than the eastern lobe. The eastern lobe is more bent and has a relatively flat spectral index compared to the western lobe. The relatively flat spectral index in the lobe is typically seen for the sources that have the effect of the cluster environment. So, it is possible that the eastern lobe of \src{} is more affected by the cluster environment. Spectral steepening in the ranges of $-1.2$ to $-2.0$ (see the left panel of Fig. \ref{fig:spec_map}) can be seen in the one-sided diffuse wing structure of J0011+3217. 

The produced spectral index map of J0011+3217 has a resolution of 21\arcsec. The radio core in the spectral index map of J0011+3217 at 21\arcsec\ resolution is indistinguishable from the surroundings. 
We measured the spectral index of the radio core of \src{} with the high-resolution images at 144 MHz, 607 MHz, and 3000 MHz in which the radio core is resolved. 
The measured spectral index of the core is $-0.33\pm 0.06,$ which is relatively flat, as expected for the core of a radio galaxy.

The nature of the spectral slope of a typical radio galaxy with jets, primary components (cores), and secondary components (hotspots) can be understood in terms of the spectral index as follows \citep{Za19}:
(1) $\alpha<-0.7$ is described as a steep spectrum and is typical for optically thin synchrotron structures, where electrons have cooled off, such as radio lobes or faint diffuse emissions in the structure. The western lobe of \src{} (in the left panel of Fig. \ref{fig:spec_map}) and the extended one-sided secondary wing of diffuse emission along the southern side of J0011+3217, all show a steep spectral index. (2) $-0.7\leq\alpha\leq-0.4$, denoted as flat, indicative of recent particle acceleration in optically thin synchrotron emission, typical for jet emission and hotspots. A similar spectral index $\sim$ --0.6 at the hotspots of both sides (eastern and western) of J0011+3217 is measured. (3)  $\alpha>-0.4$, this range of spectral index characterises sources where synchrotron self-absorption or free-free absorption is involved such as in AGN core components (primary components) as observed (fitted spectral index $\sim$ --0.33) for the core of J0011+3217.
 
The spectral index of a typical radio galaxy including the whole structure lies between $-0.7$ to $-0.8$ whereas the total spectral index as calculated for \src{} between 144 MHz and 607 MHz is $-0.97$. The steepening in the spectral index of \src{} is due to the presence of extended one-sided diffuse emission in the structure, which is as expected very steep ($-1.2$ to $-2.0$) because of the very faint diffuse emission.  

\subsection{Radio luminosity}
\label{subsec:radio_lum}
We measured the radio luminosity ($L_{\textrm{rad}}$) of this peculiar radio galaxy using the standard formula \citep{Do09}:
\begin{equation}
    L_{\textrm{rad}}=4\pi{D_{L}}^{2}S_{\nu}(1+z)^{\alpha-1}
,\end{equation}
where $\alpha$ is the spectral index, $z$ is the spectroscopic redshift of J0011+3217 ($z=0.107144$ for G1; see the right panel of Fig. \ref{fig:optical}), $D_{L}$ is the luminosity distance of J0011+3217 in metres (m), and $S_{\nu}$ denotes the flux density of J0011+3217 at a frequency $\nu$ in the unit W m$^{-2}$ Hz$^{-1}$. The total radio luminosity of J0011+3217 is calculated as 1.65$\times$ 10$^{26}$ W Hz$^{-1}$ at 144 MHz. This places it just above the nominal Fanaroff-Riley break at this frequency of $10^{26}$ W Hz$^{-1}$ and is similar to the average radio luminosity measured for 215 FRII giant radio galaxies at 144 MHz \citep{Da20a}.
 
\subsection{Black hole mass}
\label{subsec:bl-mass}
The central black hole masses of the pair of interacting galaxies (G1 and G2; see the right panel of Fig. \ref{fig:optical}) of J0011+3217 are estimated using the $M_{\text{BH}}-{\sigma}$ relation \citep{Fe00, Ge00}: 
\begin{equation}
        \frac{M_{BH}}{10^8M_\odot}=3.1 \left(\frac{\sigma_\ast}{200 \text{~km~s}^{-1}}\right)^4 
,\end{equation}
where $\sigma_{\ast}$ is the velocity dispersion, and $M_{\odot}$ is the stellar mass. Here, we use the central velocity dispersion ($\sigma_{\ast}$) for the optical hosts taken from SDSS \citep{Ahu20}. We found that the black hole masses of J001119.35+321713.8 (G1) and  J001119.75+321709.0 (G2) are 7.37$\pm$1.08 $\times$10$^8$ M$_{\odot}$ (with a velocity dispersion of 248.34$\pm${9.13} km s$^{-1}$) and 2.68$\pm$0.51 $\times$10$^8$ M$_{\odot}$ (with a velocity dispersion of 192.86$\pm${9.19} km s$^{-1}$), respectively. The optical host J001119.35+321713.8 (G1) is a luminous red galaxy and more massive than J001119.75+321709.0 (G2).

The comparison of black hole mass for both optical galaxies (G1 and G2) with statistically measured average black hole mass for radio galaxies (with the spectroscopic data collected from \citet{Be12} and giant radio galaxies by \citet{Da20b} (see Fig. 11a of \citet{Da20b}) shows that \src{} is relatively less massive.

\subsection{Projected linear size}
\label{subsec:linear_size}
We measure the projected linear size of this peculiar source along with that of the secondary wing using the following formula: 
\begin{equation}
    D_p=\frac{\theta\times D_{co}}{(1+z)}\times\frac{\pi}{648000}
,\end{equation} 
where $\theta$ is the largest angular size (LAS) of the source (in arcsec), $z$ is the redshift of J0011+3217, $D_{co}$ is the comoving distance of J0011+3217 in Mpc, and $D_p$ is the projected linear size (in Mpc). We measure the projected linear size of the primary lobe as 0.99 Mpc with the largest angular size of 407 arcsec and a comoving distance of 463.7 Mpc. The measured projected linear size of the secondary wing is 0.85 Mpc (approximately 85\% of the primary lobe), with the largest angular size of 476 arcsec. 

The cutoff projected linear size for a giant radio galaxy is usually taken as $\geq0.7$ Mpc \citep{Ku18, Da20a, Da20b, Bh24}. Therefore, based on the projected linear size of the primary lobe (0.99 Mpc) and the diffuse secondary wing (0.85 Mpc), this peculiar source is a giant radio galaxy with a giant one-sided diffuse secondary wing. 
This source is also catalogued as a giant radio galaxy in the list of 162 giant radio galaxies identified by \citet{Da20b} using the National Radio Astronomy Observatory (NRAO) VLA Sky Survey (NVSS) survey data. Because of the coarser resolution (45\arcsec) and lower sensitivity (0.45 mJy beam$^{-1}$), the NVSS survey did not detect the southward extended wing of J0011+3217 in \citet{Da20b}. The low-frequency LOFAR survey has much better sensitivity (83 $\mu$Jy beam$^{-1}$) and high resolution (6\arcsec) in comparison to NVSS, which helped us to detect the diffuse one-sided southward extended wing of J0011+3217 as presented in the current paper.

\section{Discussion}
\label{sec:discuss}
Here, we discuss the different physical characteristic features and possible formation mechanisms of the peculiar radio galaxy \src{}.

\subsection{One-sided wing behaviour}
\label{subsec:one-sided}
The peculiar radio galaxy J0011+3217 has a large diffuse secondary wing elongated out to 0.85 Mpc from the centre towards the southern part of the galaxy, which is 85\% of the size of the primary lobe. Although extremely rare, one-sided secondary lobes have previously been discovered in some XRGs. \citet{Ch07} discovered small, one-sided wings for a few X-shaped sources; however, \citet{Ro18} examined these sources using VLA at 1.4 GHz and discovered a very faint wing in the other direction for some of the sources. The source presented in the current paper also seems to have a very faint wing to the north of the eastern lobe. The faintness and small size of the wings in other directions could be a consequence of the energy losses sustained by the relativistic plasma produced in these older or fossil radio components. \citet{Bh22} also catalogued 13 sources as single-wing XRGs. The sizes of the one-sided secondary lobes in all these sources were very small (<25\% of the primary lobes). Thus, although secondary wings are also observed in X-shaped sources, giant one-sided secondary wings or plumes have never been observed in any other XRGs, as seen in the peculiar source, J0011+3217, which makes this source special. 

\subsection{Effect of jet--intergalactic medium or jet--ICM interaction}
\label{subsec:mechanism}
The presented source (J0011+3217) in this current paper possesses minor distortions or off-axis diversions (or bending of the primary lobe in a common direction) at both edges of the structure (see Fig. \ref{fig:GMRT_full}). Such bending in the jet is more common for FRI radio sources, that is, for wide-angle tailed sources \citep[WATs; e.g.][]{Ow76, Ko88, Bu98, Ma10, Min19, Mi19, Bh22, Sa22, Pa23}. WATs are typically found in dense cluster regions and form when the lobes or jets of the WAT are swept back by the relative motion of the surrounding medium that bends the jets in a common direction \citep[e.g.][]{Ow76, Ko88, Bu98, Ma10, Min19, Mi19, Pa19, Bh22, Sa22, Pa23}. 

X-ray emission can originate from radio galaxies in several ways \citep{Bi96}. Radiation identifiable with a component of a radio galaxy usually appears to come from jets, hot spots, or lobes and is consistent with either synchrotron or inverse-Compton emission \citep{Perley, Hardcastle2004, Mingo}. As shown in Fig. \ref{fig:X-ray}, no  X-ray emission is detected from the primary lobe of J0011+3217. The strong and extended thermal X-ray emission in the Abell 7 cluster may restrict the detection of faint or weak non-thermal X-ray emission from \src{} near the core and the primary lobe. As noted above, there is no evidence of thermal emission from the presumed host group of the radio galaxy either.

As discussed above, J0011+3217 is hosted by a group of galaxies that is offset from, but physically associated with, the cluster Abell 7, and it seems very likely that its unusual appearance is associated with its large-scale environment. The fact that the `missing' northern wing would point back towards the centre of the cluster is significant in this regard. We suggest that the host group is currently moving radially inwards at a speed comparable to our rough estimate of the velocity offset (i.e. around 1000 km s$^{-1}$), which means that the motion is at least mildly supersonic with respect to the cluster atmosphere, and so by definition ram pressure dominates over the static thermal pressure of the cluster. We can use the universal pressure profile of \cite{Arnaud10} to estimate that the thermal pressure of Abell 7 at $r_{500}$ is $\sim 5 \times 10^{-14}$ Pa, while the ram pressure should be of order twice this value. Using {\sc pysynch} \citep{Hardcastle98} we estimate an equipartition pressure in the active lobes around $3 \times 10^{-14}$ Pa, and we know from inverse-Compton observations that the true pressure is likely to be a factor of a few higher. The pressure in the wings must be comparable to that in the lobes since the internal sound speed of the radio galaxy is very high. The approximate equality between the ram pressure and the internal pressure of the wings means that it is highly plausible that the asymmetrical interaction with the cluster material will tend to push both the primary lobes and the secondary wings away from the cluster centre.

We propose that as a result of this interaction, the northern wing of J001+3217 has been `folded' behind the W lobe; due to projection effects, this wing is not visible for most of its length (perhaps appearing as steep-spectrum material to the N of the E lobe). It is thus possible that intrinsically J0011+3217 has both secondary wings, and that they would be visible as separate structures from a line of sight, for example through the cluster centre. Meanwhile, both buoyant effects and the large-scale motion would tend to stretch out the southern secondary wing, contributing to its very large projected size. Over the $10^8$ years needed to generate such a large radio source, the projected motion on the sky would be $\sim 100$--$200$ kpc (on the assumption of projected speeds between 1000 and 2000 km s$^{-1})$. 

This interpretation leaves open the question of what generates the wings in the first place. The close pair of interacting galaxies that form the host of J0011+3217 is certainly an environment that might be conducive to jet precession or reorientation as well as to the formation of an asymmetric small-scale hot gas environment. The multiple, offset hotspots in J0011+3217 are consistent with a jet precession model \citep{Ho23} but could also be related to the ram pressure bending of the primary lobes. Precession models for this source will be discussed in more detail in a separate paper.

\section{Conclusions}
\label{sec:conclusion}
In this paper, we have presented observations of a peculiar radio galaxy, J0011+3217, from LoTSS at 144 MHz. This source has a large one-sided secondary wing with a linear size of 0.85 Mpc (approximately 85\% of the size of the primary lobe, which has a linear size of 0.99 Mpc). The host galaxy has an interacting companion 16 kpc away in projection. The radio source is an unusual giant radio galaxy with a misaligned distorted primary lobe. J0011+3217 is located 1.2 Mpc from the centre of the cluster Abell 7. The interacting pair of optical galaxies of \src{} appears to be part of a group of galaxies that is interacting with the nearby cluster. The effect of cluster motion on the structure of \src{} is discussed in the paper.

The effect of the motion of the host group with respect to the surrounding cluster can explain the off-axis distortion of the primary lobe, the formation of the one-sided large diffuse secondary wing, and the misaligned western primary lobe of J0011+3217. To measure the detailed parameters and study the conditions under which this peculiar radio galaxy forms, further studies, including simulations and optical follow-up observations, are encouraged. Deeper X-ray observations will play a key role in our understanding of the misaligned jet or lobe interpretation and will help in the study of the environmental and clustering information.

\begin{acknowledgements}
We thank the anonymous reviewer for helpful suggestions.
This paper makes use of data of LOw-Frequency ARray (LOFAR) Two-metre Sky Survey second data release (LoTSS DR2) available at \href{https://lofar-surveys.org/}{https://lofar-surveys.org/}. LOFAR data products were provided by the LOFAR Surveys Key Science project (LSKSP; \href{ https://lofar-surveys.org/}{ https://lofar-surveys.org/}) and were derived from observations with the International LOFAR Telescope (ILT). LOFAR (van Haarlem et al. 2013) is the Low-Frequency Array designed and constructed by ASTRON. It has observing, data processing, and data storage facilities in several countries, which are owned by various parties (each with their funding sources), and which are collectively operated by the ILT foundation under a joint scientific policy. The efforts of the LSKSP have benefited from funding from the European Research Council, NOVA, NWO, CNRS-INSU, the SURF Co-operative, the UK Science and Technology Funding Council (STFC) and the Jülich Supercomputing Centre. This paper makes use of GMRT archival data available at \href{https://naps.ncra.tifr.res.in/goa/data/search}{https://naps.ncra.tifr.res.in/goa/data/search}. We thank the staff of the GMRT that made these observations possible. GMRT is run by the National Centre for Radio Astrophysics of the Tata Institute of Fundamental Research.
This publication also uses Pan-STARRS1 data available at \href{http://ps1images.stsci.edu/cgi-bin/ps1cutouts}{http://ps1images.stsci.edu/cgi-bin/ps1cutouts}. The Pan-STARRS1 Survey (PS1) has been made possible through contributions of the Institute for Astronomy, the University of Hawaii, the Pan-STARRS Project Office, the Max-Planck Society and it's participating institutes, the Max Planck Institute for Astronomy, Heidelberg and the Max Planck Institute for Extraterrestrial Physics, Garching, The Johns Hopkins University, Durham University, the University of Edinburgh, Queen's University Belfast, the Harvard-Smithsonian Center for Astrophysics, the Las Cumbres Observatory Global Telescope Network Incorporated, the National Central University of Taiwan, the Space Telescope Science Institute, the National Aeronautics and Space Administration under Grant No. NNX08AR22G issued through the Planetary Science Division of the NASA Science Mission Directorate, the National Science Foundation under Grant No. AST-1238877, the University of Maryland, and Eotvos Lorand University (ELTE).
This paper also used Sloan Digital Sky Survey V data. Funding for the Sloan Digital Sky Survey V has been provided by the Alfred P. Sloan Foundation, the Heising-Simons Foundation, the National Science Foundation, and the Participating Institutions. SDSS acknowledges support and resources from the Center for High-Performance Computing at the University of Utah. The SDSS website is \url{www.sdss.org}. SDSS is managed by the Astrophysical Research Consortium for the Participating Institutions of the SDSS Collaboration, including the Carnegie Institution for Science, Chilean National Time Allocation Committee (CNTAC) ratified researchers, the Gotham Participation Group, Harvard University, Heidelberg University, The Johns Hopkins University, L’Ecole polytechnique f\'{e}d\'{e}rale de Lausanne (EPFL), Leibniz-Institut f\"{u}r Astrophysik Potsdam (AIP), Max-Planck-Institut f\"{u}r Astronomie (MPIA Heidelberg), Max-Planck-Institut f\"{u}r Extraterrestrische Physik (MPE), Nanjing University, National Astronomical Observatories of China (NAOC), New Mexico State University, The Ohio State University, Pennsylvania State University, Smithsonian Astrophysical Observatory, Space Telescope Science Institute (STScI), the Stellar Astrophysics Participation Group, Universidad Nacional Aut\'{o}noma de M\'{e}xico, University of Arizona, University of Colorado Boulder, University of Illinois at Urbana-Champaign, University of Toronto, University of Utah, University of Virginia, Yale University, and Yunnan University. MJH acknowledges support from the STFC grant [ST/V000624/1].
 SK gratefully acknowledges the Department of Science \& Technology, Government of India for financial support, vide reference no. DST/WISE-PhD/PM/2023/3 (G) under the ‘WISE Fellowship for Ph.D.’ program to carry out this work.
\end{acknowledgements}

\begin{appendix} 
\section{Error calculations}
\label{sec:error}
\subsection{For the spectral index}
\label{sec:spec}
The uncertainty in spectral index measurements between 144 MHz and 607 MHz frequencies due to flux density uncertainty \citep{Ma16} is calculated using Eq. \ref{equ:equ2}:\begin{equation}
        \Delta\alpha=\frac{1}{\ln\frac{\nu_{1}}{\nu_{2}}}\sqrt{\left(\frac{\Delta S_{1}}{S_{1}}\right)^{2}+\left(\frac{\Delta S_{2}}{S_{2}}\right)^{2}}
        \label{equ:equ2}
,\end{equation}
where $\nu_{1, 2}$ and $S_{1, 2}$ refer to 144 MHz and 607 MHz frequencies and flux densities, respectively. The error in flux densities in LOFAR (144 MHz;  $\Delta S_{1}$) and GMRT (607 MHz) $\Delta S_{2}$ are 0.31 Jy \citep[5\% error in LOFAR image;][]{Sh22} and 0.08 Jy (calculated using Eq. \ref{eq:error}), respectively. The measured spectral index uncertainty between 144 MHz and 607 MHz using Eq. \ref{equ:equ2} is $\Delta\alpha$ = 0.05.

\subsection{For the black hole mass}
\label{sec:BH}
The uncertainty in black hole mass is measured using relative error in black hole mass ($\Delta$M$_{BH}$) and velocity dispersion ($\Delta \sigma$) as
$\Delta$M$_{BH}$ = 4$\times \Delta \sigma$, where $\Delta \sigma$ = $\frac{\delta \sigma}{\sigma}$. So the uncertainty in the black hole mass is $\delta$M$_{BH}$ = M$_{BH}$$\times$$\Delta$M$_{BH}$.
\end{appendix}


\begin{thebibliography}{}
\bibitem[{Aghanim et al.}(2020)]{Ag20}
Aghanim, N., Akrami, Y., Ashdown, M., et al. 2020, A\&A, 641, 67 

\bibitem[{Ahumada et al.}(2020)]{Ahu20}
Ahumada, R., Allende, P. C.,  Almeida, A., et al. 2020, \apjs, 249, 21 

\bibitem[{Arnaud et al.}(2005)]{Arnaud05}
Arnaud M., Pointecouteau, E., Pratt, G.W., 2005, A\&A, 441, 893

\bibitem[{Arnaud et al.}(2010)]{Arnaud10}
Arnaud M., Pratt, G.W., Piffaretti, R., et al. 2010, A\&A, 517, A92

\bibitem[{Battistini et al.}(1980)]{Ba80}
Battistini, P., Bonoli, F., Silvestro, S., et al. 1980, A\&A,, 85, 101

\bibitem[{Begelman et al.}(1979)]{Be79}
Begelman, M. C., Rees, M. J., \& Blandford, R. D. 1979, Natur, 279, 770

\bibitem[{Bera et al.}(2020)]{Be20}
Bera, S., Pal, S., Sasmal, T. K., \& Mondal, S. 2020, \apjs, 251, 9

\bibitem[{Best \& Heckman}(2012)]{Be12}
Best, P. N., \& Heckman, T. M. 2012, MNRAS, 421, 1569

\bibitem[{Birkinshaw \& Worrall}(1996)]{Bi96}
Birkinshaw M., Worrall D.M. 1996, in Hardee P.E., Bridle A.H., Zensus J.A., eds, Energy Transport in Radio Galaxies and Quasars, ASP Conference Series vol. 100, San Francisco, p. 335

\bibitem[{Botteon et al.}(2022)]{Bo22}
Botteon, A., Shimwell, T. W., \& Cassano, R. 2022, A\&A, 660, A78

\bibitem[{Bhukta et al.}(2022)]{Bh22}
Bhukta, N., Pal, S., \& Mondal, S. 2022, MNRAS, 516, 372

\bibitem[{Bhukta et al.}(2024)]{Bh24}
Bhukta, N., Manik S., Pal, S., \& Mondal, S. 2024, ApJS, in press, preprint: arXiv:2201.12353, DOI: 10.3847/1538-4365/ad5184


\bibitem[{Briggs}(1995)]{Br95}
Briggs, D. S. 1995, AAS, 27, 1444

\bibitem[{Burns}(1998)]{Bu98}
Burns, J. O. 1998, Science, 280, 400

\bibitem[{Capetti et al.}(2002)]{Ca02}
Capetti, A., Zamfir, S., Rossi, P., et al. 2002, A\&A, 394, 39

\bibitem[{Chambers et al.}(2016)]{Ch16}
Chambers, K.C., Magnier, E. A., Metcalfe, N., et al. 2019, preprint: arXiv:1612.05560

\bibitem[{Cheung}(2007)]{Ch07}
Cheung, C. C. 2007, ApJ, 133, 2097

\bibitem[{Colla et al.}(1970)]{Co70}
Colla, G., Fanti, C., Fanti, R., et al. 1970. Astr. Astrophys. Suppl, 1, 281.

\bibitem[{Cotton et al.}(2020)]{Co20}
Cotton, W.D., Thorat, K., Condon, J.J., et al. 2020, MNRAS, 495, 1271--1283
  
\bibitem[{Dabhade et al.}(2020a)]{Da20a}
Dabhade, P., Mahato, M., Bagchi, J., et al. 2020a, A\&A, 635, A5

\bibitem[{Dabhade et al.}(2020b)]{Da20b}
Dabhade, P., Mahato, M., Bagchi, J., et al. 2020b, A\&A, 642, A153

\bibitem[{den Herder et al.}(2001)]{de01}
den Herder, J. W., Brinkman, A. C., Kahn, S. M., et al. 2001, A\&A, 365, L7--L17

\bibitem[{Donoso et al.}(2009)]{Do09}
Donoso, E., Best, P. N., \& Kauffmann, G. 2009, \mnras, 392, 617

\bibitem[{Ferrarese \& Merritt}(2000)]{Fe00}
Ferrarese, L., \& Merritt, D. 2000, ApJ, 539, L9--L12

\bibitem[{Gabriel et al.}(2004)]{Ga04}
Gabriel, C., Denby, M., Fyfe, D., et al. 2004, in Astronomical Data Analysis Software and Systems XIII, eds. F. Ochsenbein, M. G. Allen, \& D. Egret, ASP Conf. Ser., 314, 759

\bibitem[{Gebhardt et al.}(2000)]{Ge00}
Gebhardt, K., Bender, R., Bower, G., et al. 2000, ApJ, 539, L13--L16

\bibitem[{Gopal-Krishna et al.}(2012)]{Go12}
Gopal-Krishna, Biermann, P. L., Gergely, L. Á \& Wiita, P. J. 2012, RAA 12, 127

\bibitem[{Hardcastle et al.}(1998)]{Hardcastle98}
Hardcastle, M.J, Birkinshaw, M., Worrall, D. M. 1998, MNRAS, 294, 615

\bibitem[{{Hardcastle} {et~al.}(2004){Hardcastle}, {Harris}, {Worrall}, \&
  {Birkinshaw}}]{Hardcastle2004}
{Hardcastle}, M.~J., {Harris}, D.~E., {Worrall}, D.~M., \& {Birkinshaw}, M. 2004, \apj, 612, 729

\bibitem[{Hardcastle {et~al.}(2023)}]{Hardcastle23}
Hardcastle, M. J., et al. 2023, A\&A, 678, 151

\bibitem[{{Harris} {et~al.}(1994){Harris}, {Carilli}, \& {Perley}}]{Perley}
{Harris}, D. E., {Carilli}, C. L., \& {Perley}, R. A. 1994, \nat, 367, 713

\bibitem[{Horton et al.}(2020)]{Ho20}
Horton, M. A., Krause, M. G. H. \& Hardcastle, M. J. 2020, MNRAS, 499, 5765

\bibitem[{Horton et al.}(2023)]{Ho23}
Horton, M. A., Krause, M. G. H. \& Hardcastle, M. J. 2023, MNRAS, 521, 2593

\bibitem[\protect\citeauthoryear{Hopkins et al.}{2003}]{phoenix} 
Hopkins, A. M., Afonso, J., Chan, B., Cram, L. E., Georgakakis, A., Mobasher, B. 2003, AJ, 125, 465

\bibitem[{Iqbal et al.}(2023)]{Iqbal23}
Iqbal, A., Pratt, G.W., Bobin, J., et al., 2023, A\&A, 679, 51

\bibitem[{Jansen et al.}(2001)]{Ja01}
Jansen, F., Lumb, D., Altieri, B., et al. 2001, A\&A, 365, L1--L6

\bibitem[{Jones et al.}(2017)]{Jo17}
Jones, T. W., Nolting, C., O’Neill, B. J., \& Mendygral, P. J. 2017, PhPl, 24, 041402

\bibitem[{Komissarov}(1988)]{Ko88}
Komissarov, S. S. 1988, Astrophysics, 29, 619

\bibitem[{Ku\'zmicz et al.}(2018)]{Ku18}
Ku\'zmicz, A., Jamrozy, M., Bronarska, K., Janda-Boczar, K., \& Saikia, D. J. 2018, ApJS, 238, 9

\bibitem[{Kumari et al.}(2022)]{Ku22}
Kumari, S., Pal, S., Bhukta, N., \& Mondal, S. K. 2022, Advances in Modern and Applied Sciences, 1, 59

\bibitem[{Lacy et al.}(2020)]{La20}
Lacy, M., Baum, S. A., Chandler, C. J., et al. 2020, PASP, 132, 035001

\bibitem[{Leahy \& Williams}(1984)]{Le84}
Leahy, J. P., \& Williams, A. G. 1984, MNRAS, 210, 929

\bibitem[{Mahony et al.}(2016)]{Ma16}
Mahony, E.K., Morganti, R., Prandoni, I., et al. 2016, MNRAS, 463, 2997

\bibitem[{Mao et al.}(2010)]{Ma10}
Mao, M. Y., Sharp, R., Saikia, D. J., et al. 2010, MNRAS, 406, 2578

\bibitem[{McMullin et al.}(2007)]{Mc07}
McMullin, J. P., Waters, B., Schiebel, D., Young, W., \& Golap, K. 2007, Astronomical Data Analysis Software and Systems XVI (ASP Conf. Ser. 376), ed. R. A. Shaw, F. Hill, \& D. J. Bell (San Francisco, CA: ASP), 127 

\bibitem[{Merritt \& Ekers}(2002)]{Me02}
Merritt, D., \& Ekers, R. D. 2002, Science, 297, 1310

\bibitem[{Missaglia et al.}(2019)]{Mi19}
Missaglia, V., Massaro, F., Capetti, A., et al. 2019, A\&A, 626, A8

\bibitem[{{Mingo} {et~al.}(2017){Mingo}, {Hardcastle}, {Ineson}, {Croston},
  {Dicken}, {Evans}, {Morganti}, \& {Tadhunter}}]{Mingo}
{Mingo}, B., {Hardcastle}, M. J., {Ineson}, J.~{Mahatma}, V., {et~al.} 2017,
  \mnras, 470, 2762

\bibitem[{Mingo et al.}(2019)]{Min19}
Mingo, B., Croston, J. H., Hardcastle, M. J., et al. 2019, MNRAS, 488, 2701
 
\bibitem[{Owen \& Ledlow}(1997)]{Ow97} 
Owen, F. N., \& Ledlow, M. J. 1997, ApJS, 108, 41

\bibitem[{Owen \& Rudnick}(1976)]{Ow76} 
Owen, F. N., \& Rudnick, L. 1976. ApJL, 205, L1

\bibitem[{Pal \& Kumari}(2023)]{Pa23}
Pal, S., \& Kumari, S. 2023, JA\&A, 44, 17

\bibitem[{Patra et al.}(2019)]{Pa19}
Patra, D., Pal, S., Konar, C., \& Chakrabarti, S. K. 2019, ApSS, 364, 72
  
\bibitem[{Perley \& Butler}(2017)]{Pe17}
Perley, R.A., \& Butler, B. J. 2017, ApJS, 230, 7

\bibitem[{Roberts et al.}(2018)]{Ro18}
Roberts, D. H., Saripalli, L., Wang, K. X., et al. 2018, ApJ, 852, 47

\bibitem[{Rozo et al.}(2018)]{Rozo15}
Rozo, E., Rykoff, E. S., Becker, M., Reddick, R. M., Weschler, R. H, 2015, MNRAS, 453, 38

\bibitem[{Saripalli \& Subrahmanyan}(2009)]{Sa09}
Saripalli, L., \& Subrahmanyan, R. 2009, ApJ, 695, 156

\bibitem[{Sasmal et al.}(2022)]{Sa22}
Sasmal, T. K., Bera, S., Pal, S., \& Mondal, S. 2022, ApJS, 259, 31 

\bibitem[{Shimwell et al.}(2022)]{Sh22}
Shimwell, T. W.,  Hardcastle, M. J., Tasse, C., et al. 2022, A\&A, 659, A1

\bibitem[{Swarup et al.}(1991)]{Sw91}
Swarup, G., Ananthkrishnan, S., Kapahi, V. K., Rao, A. P., Subrahamanya, C. R., \& Kulkarni, V. K. 1991, Current Science, 60, 90

\bibitem[{Worrall et al.}(1995)]{Wo95}
Worrall, D. M., Birkinshaw, M.,\& Cameron, R. A. 1995, ApJ, 449, 93

\bibitem[{Yang et al.}(2019)]{Ya19} 
Yang, X., Joshi, R., Gopal-Krishna., et al. 2019, ApJS, 245, 17

\bibitem[{Zajaček et al.}(2019)]{Za19}
Zajaček, M., Busch, G.,  Valencia-S, M., et al. 2019, A\&A, 630, A83

\end{thebibliography}
\end{document}